\documentclass[aps,pra,preprint,showpacs,groupedaddress,superscriptaddress,longbibliography]{revtex4-1}

\usepackage[english]{babel}
\usepackage{float}
\usepackage{graphicx}
\usepackage[caption=false]{subfig}
\usepackage{amsmath}
\usepackage{amsfonts}
\usepackage{amssymb}
\usepackage{mathtools}
\usepackage{amsthm,enumitem}

\begin{document}

\title{Superexponential Self-Interacting Oscillator}

 \author{Peter Schmelcher}
  \email{Peter.Schmelcher@physnet.uni-hamburg.de}
 \affiliation{Zentrum f\"ur Optische Quantentechnologien, Universit\"at Hamburg, Luruper Chaussee 149, 22761 Hamburg, Germany}
 \affiliation{The Hamburg Centre for Ultrafast Imaging, Universit\"at Hamburg, Luruper Chaussee 149, 22761 Hamburg, Germany}

\date{\today}

\begin{abstract}
The superexponential self-interacting oscillator (SSO) is introduced and analyzed. 
Its power law potential is characterized by the dependence of both the base and the exponent
on the dynamical variable of the oscillator.
Opposite to standard oscillators such as the (an-)harmonic oscillator the SSO combines both scattering and confined
periodic motion with an exponentially varying nonlinearity. The SSO potential exhibits
a transition point with a hierarchy of singularities of logarithmic and power law character
leaving their fingerprints in the agglomeration of its phase space curves. The period of the
SSO consequently undergoes a crossover from decreasing linear to a nonlinearly increasing
behaviour when passing the transition energy. We explore its dynamics and show that the 
crossover involves a kick-like behaviour. A symmetric double well variant of the SSO is
briefly discussed.
\end{abstract}

\maketitle

\section{Introduction} 
\label{sec:introduction}

Nonlinearity and related nonlinear phenomena are to a large extent responsible for the
enormous variety of complex behaviour we observe in nature. In nonlinear Hamiltonian systems the
transition from regularity to chaos \cite{Tabor,Strogatz,Reichl} in few- and many-body systems exhibits
an very rich phase space. The latter includes not only islands of (quasi-)periodic motion
in a chaotic 'sea' but also self-similar and fractal structures leading to stickiness, Levy-flights and the
transition from normal to anomalous diffusion \cite{Reichl,Altmann}. Breaking certain spatio-temporal symmetries
leads in the presence of nonlinearity or more specifically a mixed phase space, to directed transport and currents in
extended (lattice) systems although the average applied force vanishes \cite{Schanz}.
Including dissipation and noise a plethora of (strange) attractors can be observed in driven lattices.
They can be used e.g. for the simultaneous control of multi-species particle transport and their segregation in
driven lattices by their physical properties like mass, friction or size \cite{Mukhopadhyay}.
Turning to continuous media and nonlinear wave equations nonlinearity is responsible for the
existence of a zoo of nonlinear excitations, such as non-dispersive wave packets or solitons and
vortices (see e.g. \cite{Kevrekidis} for such nonlinear waves in Bose-Einstein condensates).

The most fundamental and unique building blocks which serve as model systems for our comprehensive
understanding of the structure and dynamics of more complex systems are oscillators. Covering the 
route from simplicity to complexity and from single to many-body systems often corresponds to increasing 
the number of coupled oscillators thereby providing invaluable insights into the relevant structures and dynamical
mechanisms of more realistic systems occuring in nature. While systems of linearly coupled harmonic
oscillators exhibit by construction a regular phase space adding nonlinearity leads to a rapid
transition from regularity to a mixed phase space and for strong anharmonicities to a dominant chaotic 
behaviour. Adding forcing, resonance phenomena are ubiquitous and have been used also in combination 
with dissipation to control the response of the oscillators dynamics (see references
\cite{Weinstock,Kovacic,Korsch,Reichl,Stoeckmann} addressing a variety of nonlinear oscillators
such as the Duffing oscillator, the parametrically driven oscillator or the kicked rotor). 
Specifically for the case of a parametric driving the (harmonic) oscillator can gain energy
for certain frequency ratios of the natural frequency compared to the parametric driving frequency
accompanied by a phase-locking to the parametric variation \cite{Fossen}.

Recently, the need for a deeper understanding of the mechanisms of exponential acceleration
as well as the search for its most fundamental building blocks emerged
\cite{Turaev1,Shah1,Liebchen,Shah2,Turaev2,Turaev3,Batistic}.
In this context the driven power law oscillator (TPO) \cite{Schmelcher} has been introduced and explored.
Opposite to the above-mentioned harmonic and anharmonic time-dependent oscillators
the TPO is characterized by a periodically time-dependent exponent of the 
potential such that the confinement changes its shape periodically in time covering a 
continuous spectrum of anharmonicities in a single cycle of the driving.
This allows for a two component phase space consisting not only of mixed regular and
chaotic bounded motion but in particular also of unbounded motion and exponential net
growth of the corresponding energies. Phases of motion with huge energy gain and loss
then alternate during a single driving period of the oscillator. This way it was
demonstrated \cite{Schmelcher} that driving the power of the oscillator instead of the traditional
approach of driving a parameter, such as the amplitude of an external field, provides a 
very rich phenomenology.

In this work, we go one step further along the above lines
and render the exponent dynamical in the sense that it becomes proportional to the oscillator
coordinate $q$ itself. This superexponential self-interacting oscillator (SSO) represents a conservative
Hamiltonian system with an overwhelmingly simple appearance but, as we shall see below,
with interesting (phase space) properties. While standard potential functions for oscillators
are of power law form with some given exponent, be it short- or long-ranged,
the present oscillator puts the nonlinearity to its extreme in the sense of the potential
$V(q)=|q|^q$-form describing a highly nonlinear self-interacting or self-coupled oscillator.
As a consequence our basic intuition of what an oscillator could deliver to us
and what its phenomenology and behaviour could be, have to be carefully rethought
when considering the SSO with its uncommon properties. Our aim is to develop a first understanding
of this SSO by performing a computational study of its structure and dynamical behaviour.
We remark already here, that a direct extended analytical approach is, in spite of the fact that we
are dealing with a one-dimensional integrable system, hindered by the exponential form of $V$
which prevents closed form analytical solutions of the quantities of interest (see below).
Specifically we will show that the phase space curves agglomerate at $q=0$ which is reflected in
the corresponding analytical structure of $V$ whose derivatives exhibit an intriguing structure
such that with increasing order of the derivative at $q=0$ an increasingly singular behaviour
occurs. This leads, among others, to a transition in the behaviour of the period of the
oscillator with increasing energy: for low energies it decreases linearly and once it reaches
the transition point $q=0$ a crossover to a highly nonlinear increase is observed.
This transition leaves its fingerprints also in the dynamics: at the transition point the
oscillator dynamics experiences a kick being qualitatively different for low and high energies.

In detail we proceed as follows. In section II we introduce and discuss the SSO and
derive its phase space properties. Section III contains a discussion of the analytical structure
of the potential thereby demonstrating the hierarchy of the singularities of its derivatives.
A frequency analysis of the SSO with varying energy is presented in section IV and compared
to several well-known cases. Section V contains a discussion of the dynamics showing qualitatively
different behaviour for the trajectories. A symmetric variant of our SSO is presented and analyzed in section VI.
Finally section VII provides some brief conclusions and an outlook.

\section{The SSO Hamiltonian and Its Phase Space} \label{sec:setup}

\begin{figure}
\includegraphics[width=8cm,height=8cm]{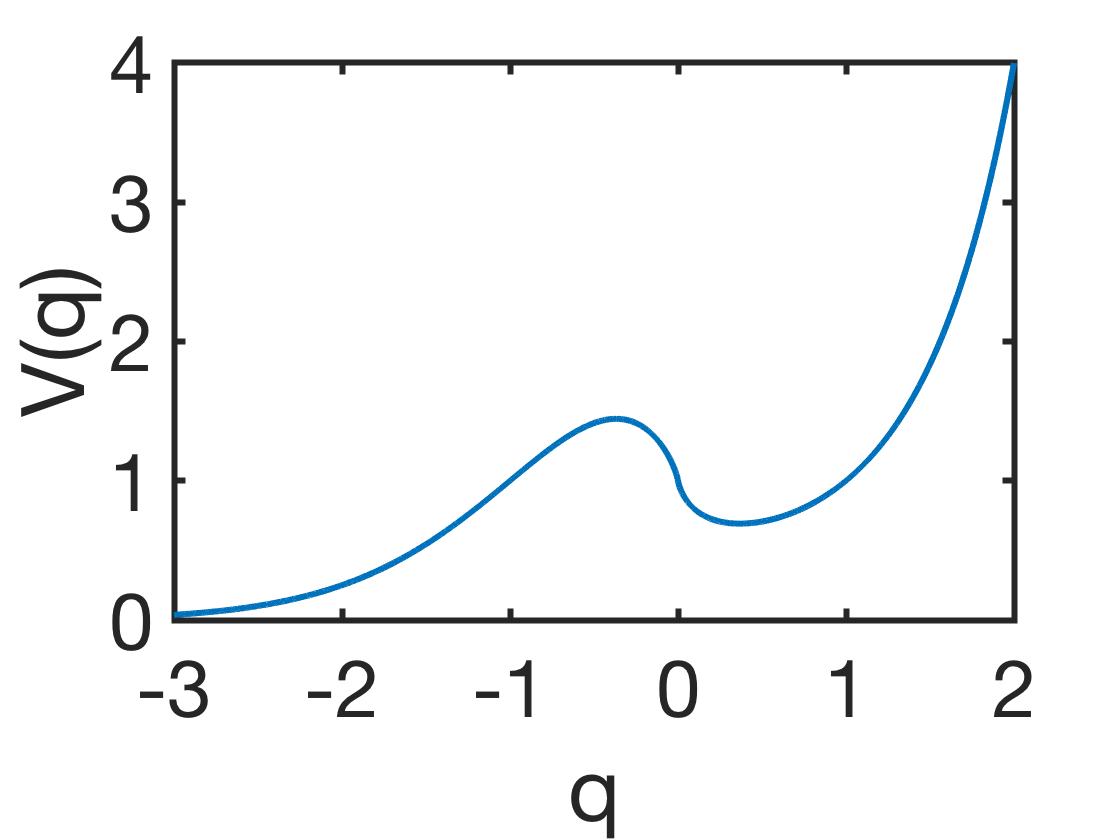} 
\caption{\label{fig:fig1} The potential $V(q)=|q|^q$ of the SSO as a function of the oscillator coordinate $q$.}
\end{figure}

Power law potentials $V(q)=q^{\alpha}$ with a constant $\alpha = 2n, n \in \mathbb{N}$ covering the harmonic
($n=1$), quartic anharmonic ($n=2$) and arbitrarily higher order cases lead to some frequently
used oscillator models. Here we relax the requirement of having a constant power and even render
it dynamical i.e. we choose it proportional to the continuously varying coordinate $q$ of the oscillator
itself. This way we end up with an extreme nonlinearity which changes with time in the corresponding
dynamics and our potential reads $V(q)=|q|^q$ occuring in the oscillator Hamiltonian 
$H=\frac{p^2}{2} + V(q)$ which, of course, conserves energy. The absolute value is
chosen to avoid complex values of the potential and the related ambiguities. The SSO potential can be
seen as if the oscillator interacts with itself in some effective manner. The peculiar shape of $V(q)$
is shown in figure \ref{fig:fig1}. It is of highly nonlinear and asymmetric character with a
confining wall for $q \rightarrow + \infty$ and approaches asymptotically a constant
(value zero) for $q \rightarrow - \infty$. In between a potential well develops. In spite of its
very simple appearance the SSO therefore accomodates both confined and bounded periodic motion
as well as scattering dynamics.

Let us discuss the potential of the SSO in some more detail due to its peculiar shape and its
varying nonlinear behaviour. For $q > 1$ the SSO 'feels' with continuously increasing $q$ a continuously
increasing stronger confinement and corresponding restorative force, or, in other words,
it encounters a rising potential of exponentially increasing steepness. Both
the potential and its 'steepness' diverge (see section \ref{sec:pot} for an analysis of its analytical
structure) for $q \rightarrow \infty$. With decreasing $q$ for $q < 1$ the SSO covers continuously
all decreasing powers, i.e. roots, which leads to a decrease of the slope of $V(q)$ until it reaches
a vanishing slope at the minimum $q_{min}=e^{-1}$ with the value 
$V(q_{min})= (\frac{1}{e})^{\frac{1}{e}}=e^{-\frac{1}{e}}$. For a further decreasing $q$ in the interval
$0<q<q_{min}$ the decreasing exponent and decreasing base leads to an increase of $V(q)=|q|^q$ until
$q$ reaches zero where $V(q)$ reaches the value one. This specific point is of major importance
for the overall behaviour and transitions of the SSO, as we shall discuss in the following sections.
In this region the slope of $V$ decreases until it becomes $-\infty$ at $q=0$
(this leads to a kick-like dynamics at this point, see section \ref{sec:dyn}). For $q_{max}<q<0$ and
decreasing $q$ the slope of $V$ increases again until it reaches zero at the maximum $q_{max}=-e^{-1}$
of the potential where its value is $V(q_{max})= (\frac{1}{e})^{-\frac{1}{e}}=e^{\frac{1}{e}}$.
Therefore, the minimum and maximum of $V$ are placed symmetrically around zero.
Finally for $q<q_{max}$, where the potential $V$ is purely repulsive towards $-\infty$, the slope of
it first increases, reaches a maximum and then decreases again until it reaches asymptotically
for $q \rightarrow -\infty$ the value zero. We note that introducing additional prefactors leading to
$V(q)= \alpha |q|^{\beta q}$ allows for a squeezing and widening of the potential well via $\beta$
whereas $\alpha$ simply scales the overall values.

\begin{figure}
\hspace*{2cm} \includegraphics[width=19cm,height=10cm]{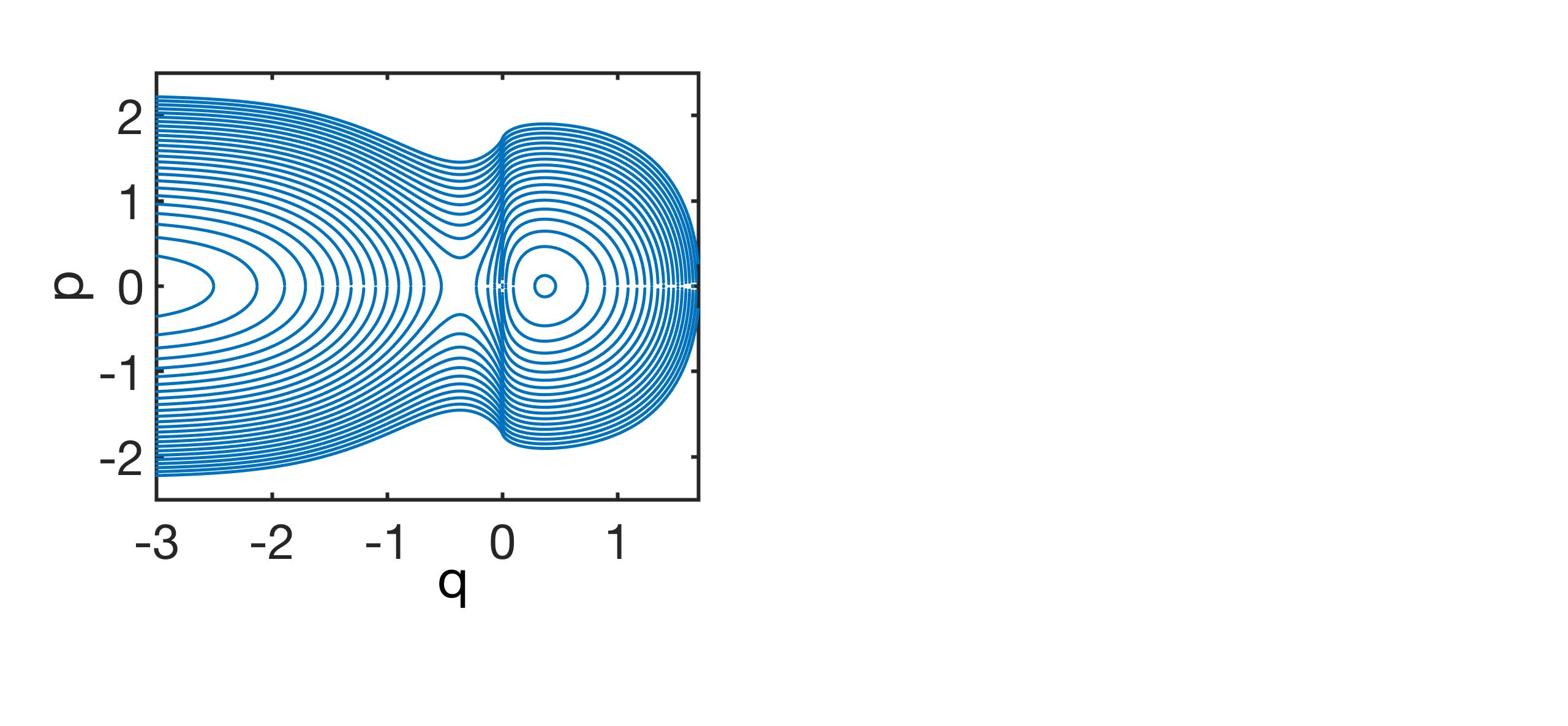} \vspace*{-3cm}
\caption{\label{fig:fig2} Phase space $(q,p)$ of the SSO. Agglomerating phase space curves
are clearly visible in the vicinity of $q=0$.}
\end{figure}

The phase space of the SSO is given by the set of curves $(q,p=\pm \sqrt{2(E-|q|^q)})$,
as shown in figure \ref{fig:fig2}. It exhibits, as expected, an elliptic
fixed point at $(q_{min},p=0)$ with surrounding elliptical curves that correspond to periodic motion
in the potential well of $V(q)$ (see figure \ref{fig:fig1}). The saddle point at $(q_{max},p=0)$ separates
the confined motion in this well and the left-incoming ($q= -\infty$) scattering trajectories
from the unbounded motion that is equally left-incoming and orbits the potential well i.e. the latter
trajectories are scattered by the well above its confining energies. 

The striking feature of the
phase space decomposition is, however, the agglomeration of the curves at $q=0$ which can be clearly
observed in figure \ref{fig:fig2}. To quantify this behaviour let us define the distance $S$
between two neighboring phase space curves ($q_1,p_1$) and ($q_2,p_2$) as 

\begin{equation}
S = \sqrt{ \left( q_2-q_{min} \right)^2 + {p_2}^2 } - \sqrt{ \left( q_1-q_{min} \right)^2 + {p_1}^2 } 
\end{equation}

which varies with the angle $\Phi$ defined as

\begin{equation}
\Phi = \text{arctan} \left( \frac{p}{q-q_{min}} \right) 
\end{equation}

\begin{figure}
\hspace*{2cm} \includegraphics[width=19cm,height=10cm]{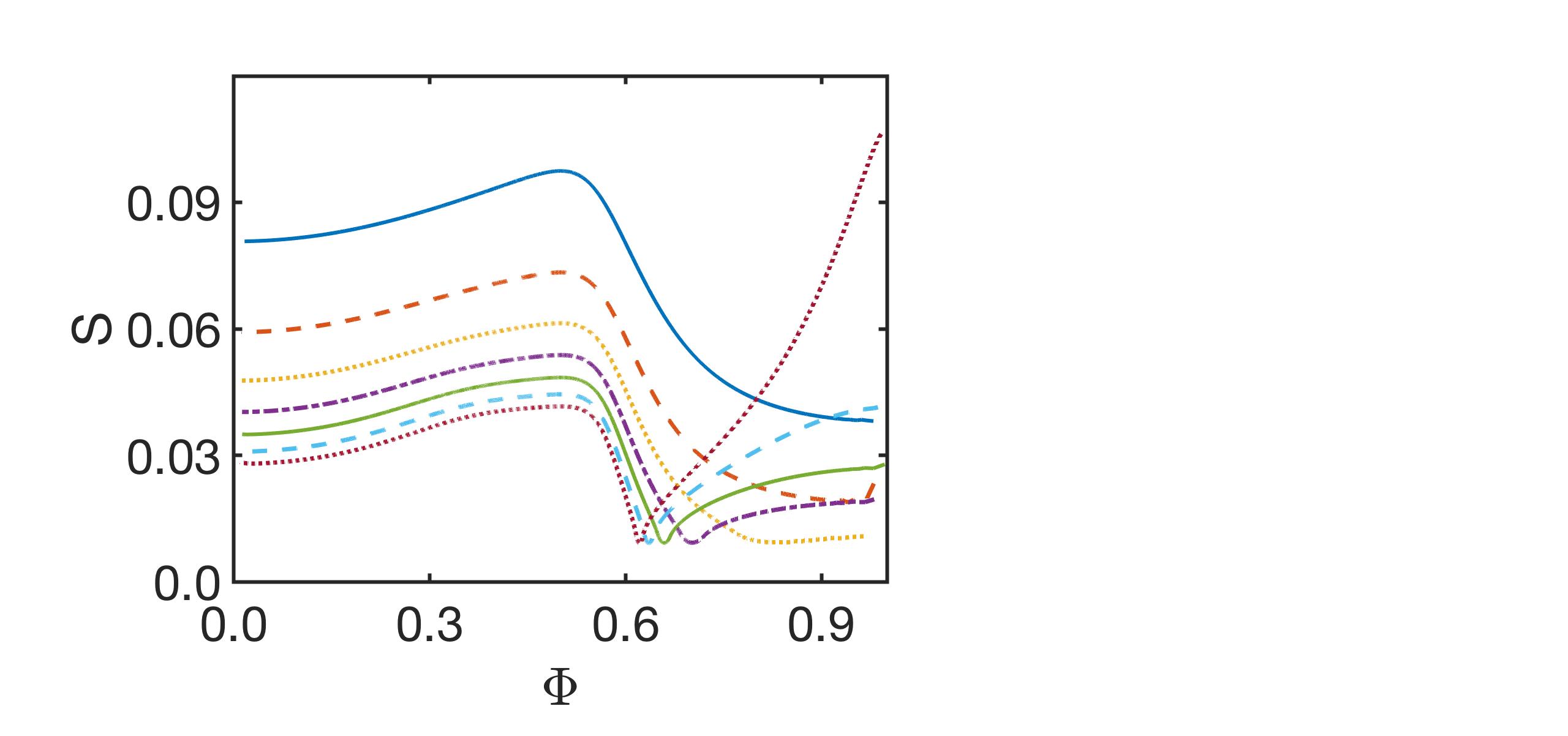} \vspace*{-1.5cm}
\caption{\label{fig:fig3} Distance $S$ of phase space curves with varying angle $\Phi$. Pronounced
minima develop for energies above $E=1$. Curves shown are for (from top to bottom) energies
$E=0.8,0.9,1.0,1.1,1.2,1.3,1.39$ and their energetically distant partner separated by the energy
$0.05$. The angle $\Phi$ is measured with respect to the elliptical fixed point ($q_{min},p=0$) and the axis $p=0$
in units of $2 \pi$.}
\end{figure}

which means that it is the azimuthal angle with respect to the axis $p=0$ for the coordinate
system centered at the elliptical fixed point ($q_{min},0$). Figure \ref{fig:fig3} shows this
distance $S (\Phi)$ for a series of different neighboring phase space
curves with equal energetical separation at $\Phi =0$ (the reader should note that the distance $S$
is the Euclidian distance in phase space and varies with $\Phi$ whereas, of course,
the energetical separation of the curves w.r.t. $H$ is constant with varying $\Phi$).

For energies below $E=1$ we observe no minima of $S (\Phi)$ within the relevant region 
$0< \Phi <1$. For $E > 1$, however, a minimum and increasingly pronounced dip develops which indicates
the point of agglomeration of the phase space curves at $q=0$ visible in figure \ref{fig:fig2}.

\section{Analytical Structure of the SSO Potential} \label{sec:pot}

It is very instructive to inspect the analytical structure of the potential $V(q) = |q|^q$, since
this will be a basis of the analysis and understanding of the dynamics and properties of the SSO in the following
(see sections \ref{sec:freq} and \ref{sec:dyn}). We do this
step by step, first addressing some lower order derivatives and subsequently providing the general expression
for the $N-$th order derivative. The first two derivatives read as follows

\begin{eqnarray}
V^{(1)}(q) = |q|^q \left( \text{ln}|q| + 1 \right) \\
V^{(2)}(q) = |q|^q \left( \left( \text{ln}|q| + 1 \right)^2 + \frac{1}{q} \right) \\ \nonumber
\end{eqnarray}

\begin{figure}
\includegraphics[width=9cm,height=8cm]{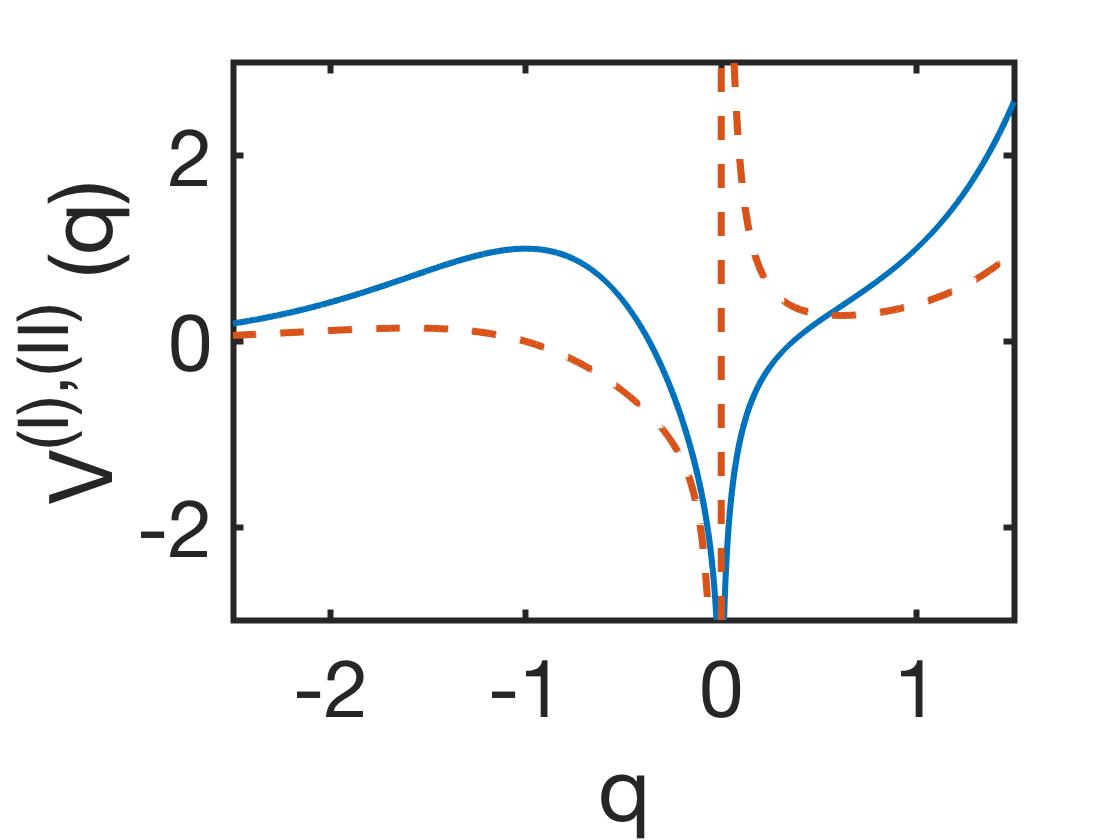} \vspace*{-0.5cm}
\caption{\label{fig:fig4} First $V^{(1)}$ (full line) and second $V^{(2)}$ (dashed line)
derivative of the potential $V$ as a function of the coordinate $q$. For reasons of illustration
the second derivative is scaled by a factor of $0.2$.}
\end{figure}

They are illustrated in Figure \ref{fig:fig4}.
$V^{(1)} (q)$ possesses a logarithmic singularity at $q=0$ with a negative infinite slope
being of the same kind for both $q \rightarrow 0^{\pm}$. $V^{(2)} (q)$ possesses both
a logarithmic and a power law singularity with exponent $-1$ at $q=0$. The latter dominates
the behaviour for $q \rightarrow 0^{\pm}$ and leads to the corresponding antisymmetry (see 
Figure \ref{fig:fig4}). Note the highly nonlinear and nonmonotonous behaviour of the derivatives,
where, in certain regions, the second derivative is even larger than the first derivative.

This translates to the $N$-th derivative which is given by

\begin{equation}
V^{(N)} \left( q \right) = {\cal{A}}^{(1)} \sum_{\substack{\{n_i \}|{\cal{C}}_{N}\\ i=1,...,N}}
\alpha \left( n_1,...,n_{N} \right) \prod_{j=1}^{N} \left({\cal{B}}^{(j)}\right)^{n_j}
\end{equation}

where ${\cal{A}}^{(1)} = |q|^q$ and

\begin{eqnarray}
\begin{array}{ll}
{\cal{B}}^{(1)} = \left( \text{ln}(|q|) +1 \right), \hspace*{0.2cm} & {\cal{B}}^{(j)} =
\left( j-2 \right) ! \left( -1 \right)^j \frac{1}{q^{j-1}}, \hspace*{0.2cm}
\text{for} \hspace*{0.1cm} j \geq 2 \\ \nonumber
\end{array}
\end{eqnarray}

for ${\cal{C}}_{N} = \sum_{i=1}^{N} i n_i = N$ and the recursion

\begin{equation}
\alpha \left(n_1,n_2,....,n_{N} \right) = \alpha \left(n_1-1,n_2,....,n_{N},0 \right)
+ \sum_{j=1}^{N} \left( n_j +1 \right) \alpha \left(n_1,...,n_{j}+1, n_{j}-1,...,n_{N} \right)
\end{equation}

holds with $\alpha(N,0,...,0) = \alpha(0,...,0,1)=1$ and negative entries are excluded.
The $N$-th derivative $V^{(N)} (q)$ is therefore a product of the original function $|q|^q$ and 
a $N$-th order polynomial consisting of powers of logarithms (${\cal{B}}^{(1)}$) and inverse powers (${\cal{B}}^{(j)}, j \ge 2$).
The dominant singularity for $q \rightarrow 0^{\pm}$ is therefore of the order of the highest inverse power 
$\propto \frac{1}{q^{N-1}}$ reflecting the symmetry in the vicinity of $q=0$.

To conclude this section, let us briefly discuss the Taylor expansion in terms of powers around the minimum $q_{min}=e^{-1}$ of
the potential well of the SSO. Obviously all ${\cal{B}}^{(1)}$-involving terms of the $N$-th derivative vanish.
The second, third and fourth derivative at $q_{min}$ amount to $e \beta, -e^2 \beta, (3 e^2 + 2 e^3) \beta$, respectively,
and the $N$-th derivative dominant term scales as $\propto e^{N-1}$ which shows their growth with increasing order. 
The equal presence of odd derivatives in the expansion is indicative of the strong (reflection) asymmetry of the potential
well around its minimum. 

\section{Frequency Analysis of the SSO} \label{sec:freq}

\begin{figure}
\hspace*{2cm}
\includegraphics[width=20cm,height=8cm]{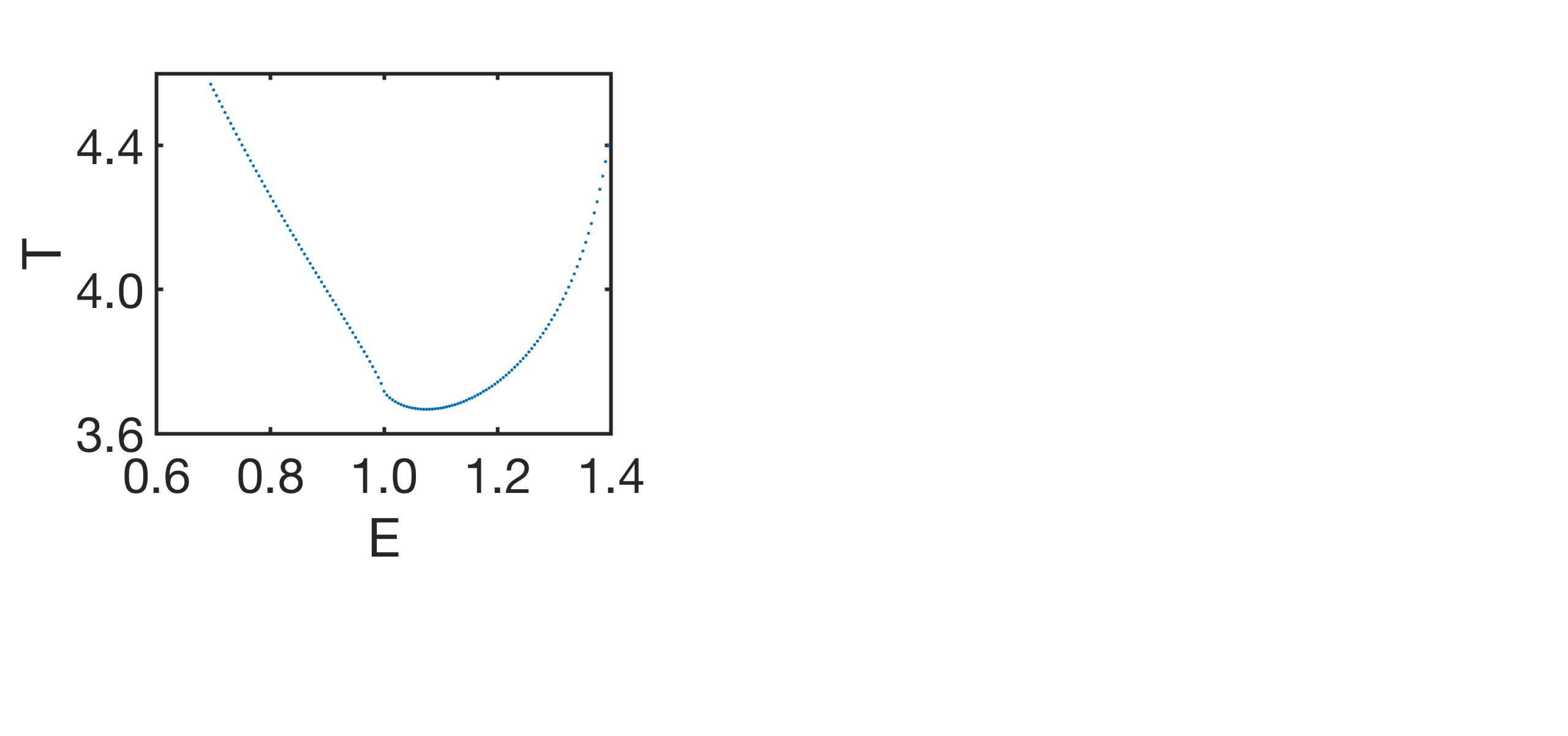} \vspace*{-3cm}
\caption{\label{fig:fig5} The period $T$ as a function of the energy $E$ covering the regime from the minimum
to the saddlepoint of the potential $V(q) = |q|^q$.}
\end{figure}

An important characteristic of oscillators is the energy dependence of their period or frequency. In the following
we analyze this dependence for our SSO. Before doing so, however, let us remind the reader of some facts
concerning the periods of some standard oscillators. Given a potential $V(q)$ the period as a function
of the energy reads 

\begin{equation}
T(E) = 2 \int \limits_{-q_0}^{q_0} \left(2 \left(E - V(q) \right) \right)^{-\frac{1}{2}} dq 
\end{equation}

where $\pm q_0$ represent the turning points of the one-dimensional periodic motion.
For the simple harmonic oscillator $V(q)= \frac{1}{2} \alpha q^2$ we have $T=\frac{2}{\sqrt{\alpha}} \pi$
which is independent of the energy. For a general power law oscillator $V(q) = \gamma q^n$
one obtains

\begin{equation}
T \left(E,\gamma \right) = \sqrt{\frac{2}{E}} \left(\frac{E}{\gamma}\right)^{\frac{1}{2n}} \int \limits_{-1}^{+1}
\left(1 - q^{2n} \right)^{-\frac{1}{2}} dq
\end{equation}

which demonstrates that the period always decreases nonlinearly with the energy. In the extreme limit $n \rightarrow \infty$
of a box potential the ballistic periodic motion yields $T \propto E^{-\frac{1}{2}}$ which is the strongest possible 
'decay' with increasing energy for power law oscillators of the type mentioned above. Figure \ref{fig:fig5}
shows $T(E)$ for the SSO which exhibits a qualitatively very different behaviour. Indeed the first property one
observes is that it consists of two components. For energies $E < 1$ which is the regime below the crossover
point $E=1$ (see also section \ref{sec:pot} for a discussion of the analytical behaviour of the SSO potential
at $E=1$ corresponding to $q=0$) $T(E)$ of the SSO shows a linear decrease within the numerical accuracy.
This is, as compared to the above-mentioned power law oscillators, a very unusual functional dependence of
the period on the energy. At $E=1$ a crossover takes place to a highly nonlinear behaviour of $T(E)$. It first
decreases, thereby showing a minimum, and consequently increases steeply. Therefore, the SSO exhibits, in spite
of its simple appearance and integrability, a stunningly complex behaviour with a crossover or transition point mediating between
two regimes of qualitatively different behaviour. It is the singular behaviour of the SSO-potential which allows
for this transition. We note here, that closed form analytical expressions are not possible due to the appearance
of the coordinate $q$ in both the basis and exponent of the potential $V$ of the SSO. 

\section{Dynamics of the SSO} \label{sec:dyn}

\begin{figure}
\hspace*{2cm}
\includegraphics[width=20cm,height=8cm]{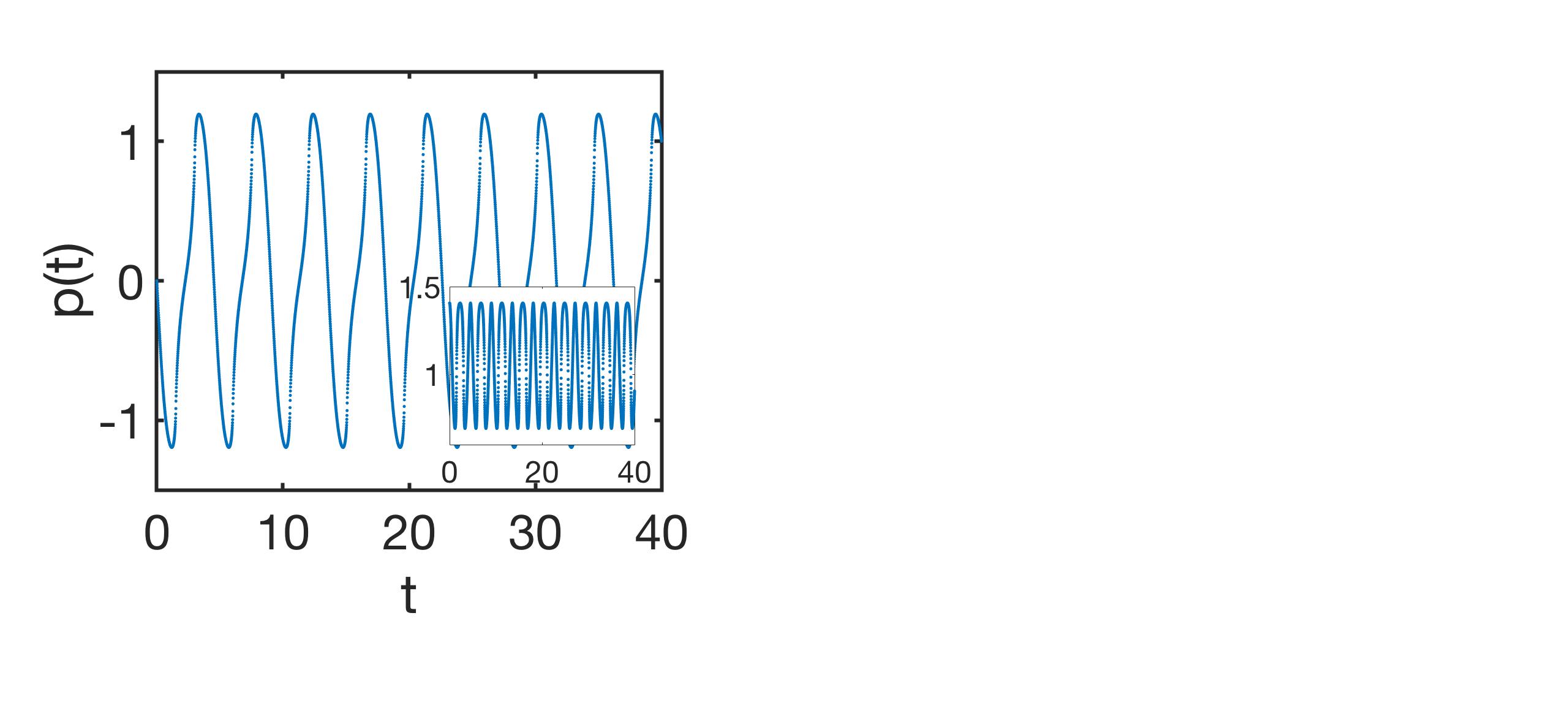} \vspace*{-2cm}
\caption{\label{fig:fig6} Time evolution of the momentum $p(t)$ and of the potential energy $E_{pot}(t)$ (inset)
for a trajectory with initial conditions $q(t=0)=1.3,~p(t=0)=0$.}
\end{figure}

This section is dedicated to the dynamics of the SSO which is governed by the Hamiltonian equations of motion

\begin{equation}
\dot{q} = p \hspace*{0.5cm} \text{and} \hspace*{0.5cm} \dot{p} = - \left( q^2 + \epsilon \right)^{\frac{q}{2}} 
\left( \text{ln} \left( \left(q^2 + \epsilon \right)^{\frac{1}{2}} \right) + \frac{q^2}{q^2 + \epsilon} \right)
\end{equation}

where a cutoff parameter $\epsilon$ has been introduced to regularize the singularities. Let us analyze the 
motion within the confining well of the SSO potential. Figure \ref{fig:fig6} shows the momentum $p(t)$
of a typical trajectory in the well for an energy $E=1.3^{1.3} \approx 1.4$ above the transition point ($E=1$)
together with its potential energy $E_{pot}(t)$. In the course of the periodic motion a certain asymmmetry is 
clearly observable. In the half-period of the motion from maximally positive to maximally negative momentum the trajectory
resides on the right half of the well which is governed by large values of $q$ and therefore also of the power
of the potential. This leads to a large force and a rapid change of momentum as compared to the motion
in the other half-period which takes place in the left half of the well. It is therefore the pronounced
asymmetry of $V(q)=|q|^q$ around its minimum which leads to this asymmetric cycling of the trajectory and
correspondingly also of its potential energy (see inset of Figure \ref{fig:fig6}). 

\begin{figure}
\hspace*{2cm}
\includegraphics[width=20cm,height=8cm]{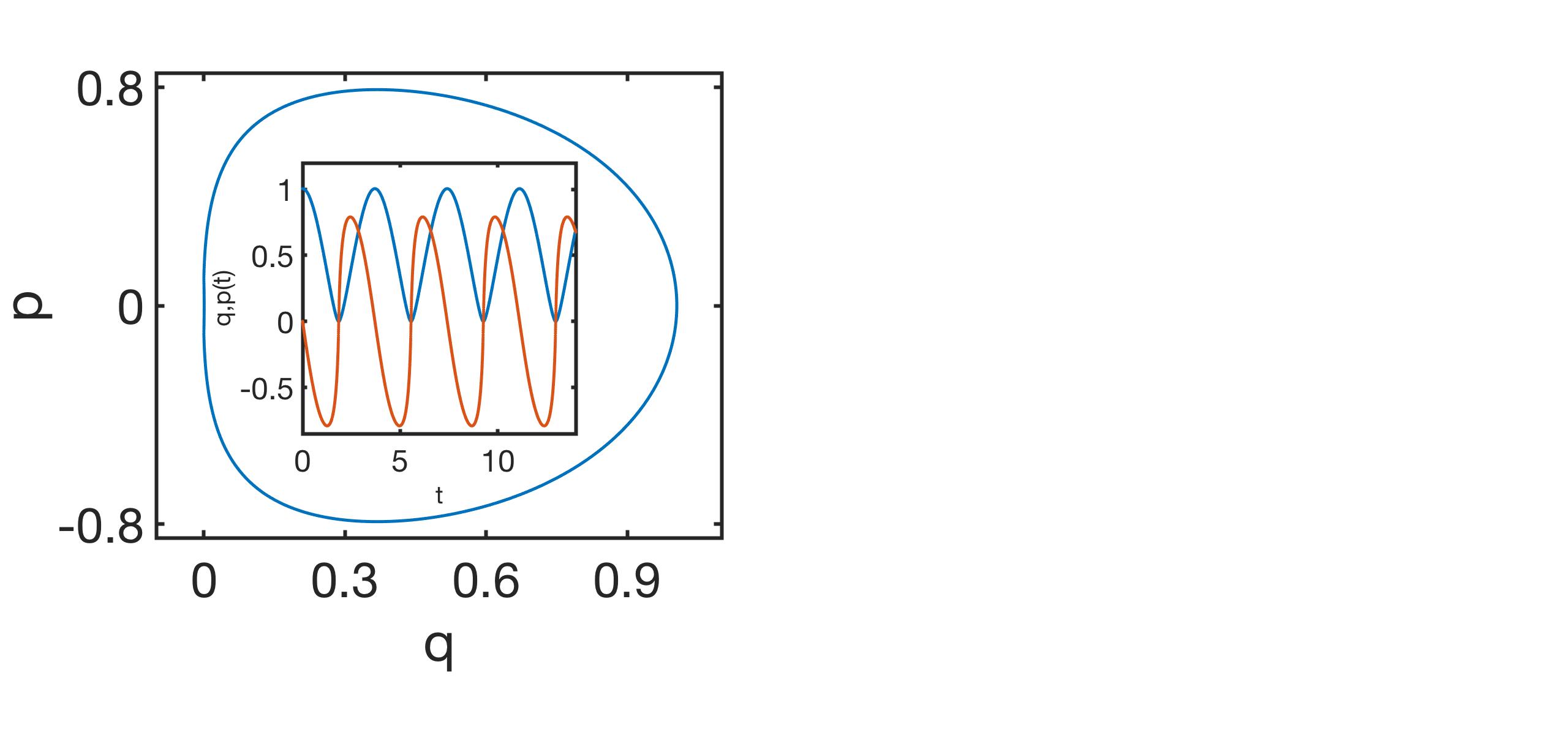} \vspace*{-2cm}
\caption{\label{fig:fig7} Phase space curve $(q,p)$ for initial conditions $q=1.005,p=0$ - observing the
kick at $q=0$ - together with the corresponding inset showing separately $q(t)$ (upper curve) and $p(t)$
(lower curve).}
\end{figure}

However, it is not only the asymmetry of the potential well, and more generally of $V(q)$
which leads to a distinct behaviour in the dynamics of the trajectories,
it is also the transition point $q=0$ that leaves its imprints. The latter we have already observed in
section \ref{sec:setup} where we analyzed the phase space: Phase space curves colaesce at the transition point,
i.e. they approach each other. This leads, as shown in Figure \ref{fig:fig7}, to an almost vertical phase
space curve particularly for energies close to the transition point $q=0$.
This means that an almost instantaneous, i.e. kick-like, momentum transfer takes place.
The latter is confirmed by the corresponding behaviour of the
coordinate $q(t)$ and momentum $p(t)$, see the inset of Figure \ref{fig:fig7}. 
$q(t)$ shows a kink-like dynamics at $q=0$ compared to an otherwise smoothly changing
behaviour. $p(t)$ exhibits a steep slope at the kinks of $q(t)$ demonstrating a large momentum
transfer whereas it is otherwise varying much more slowly. For energies off the transition point
$E=1$ the above-discussed behaviour is smoothened out.

\section{A symmetric variant of the SSO} \label{sec:var}

\begin{figure}
\hspace*{2cm}
\includegraphics[width=20cm,height=8cm]{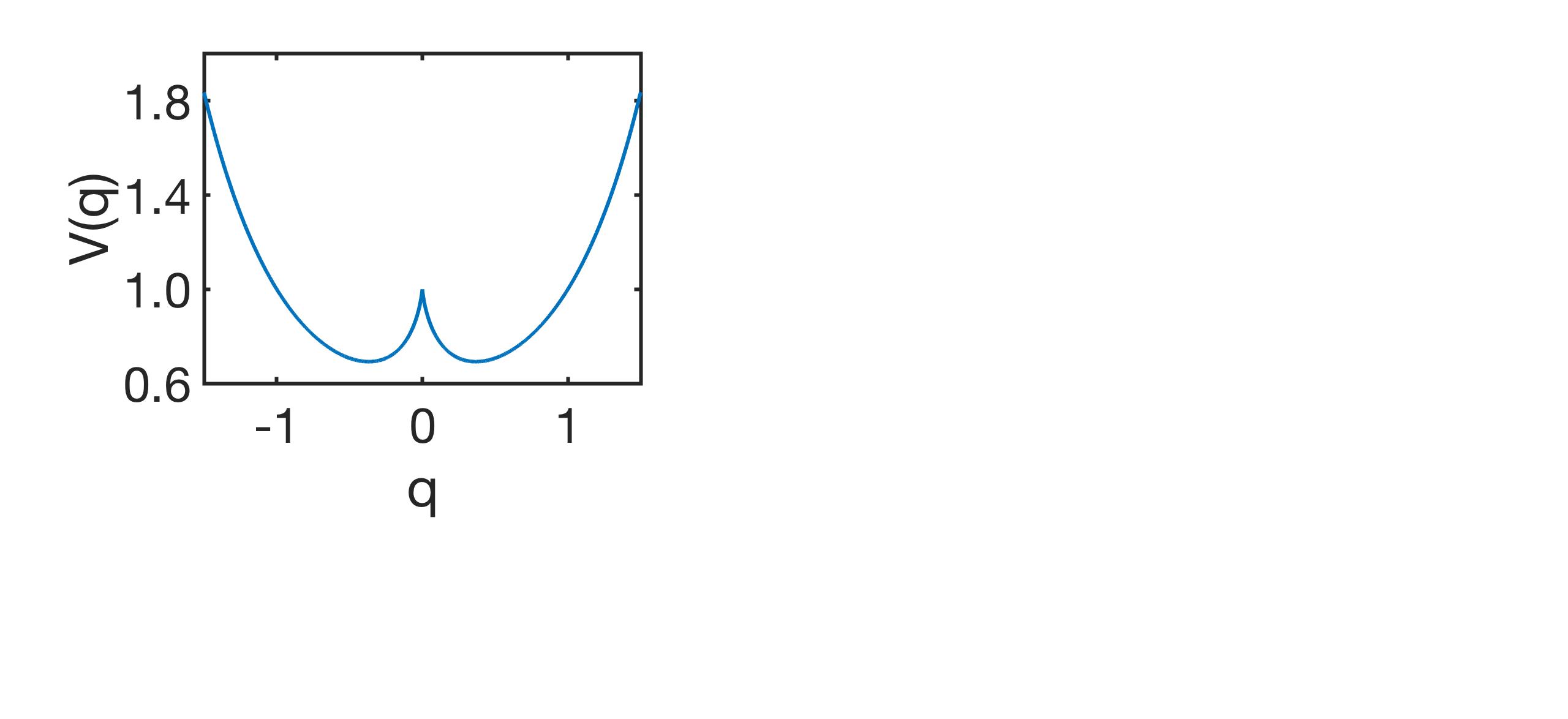} \vspace*{-2cm}
\caption{\label{fig:fig8} The potential $V(q)=|q|^{|q|}$ obtained by symmetrizing the potential
of the SSO: A double well structure emerges with a kink at the origin.}
\end{figure}

While the SSO shows some unique and interesting properties, in particular in comparison with known
oscillators of equally simple appearance, it is by no means the only superexponential and even
self-interacting oscillator possible. One immediate variant of the SSO is obtained if one considers
the symmetrized version of it (SVO) which takes on the following appearance

\begin{equation}
H=\frac{p^2}{2} + |q|^{|q|}
\end{equation}

which is obtained from the SSO Hamiltonian by taking the absolute value of the corresponding
exponent of the potential. Its potential is shown in Figure \ref{fig:fig8} where the reflection
symmetry around $q=0$ is obvious. As a consequence of taking the absolute value of the exponent
we encounter now a double well with exclusively confined motion and two symmetrically
placed wells and minima at $q_{min}= \pm \frac{1}{e}$ but no scattering trajectories
as in the case of the SSO. The barrier of the double well possesses a kink at the origin 
which represents its maximum. Indeed, the potential of the SVO is composed of two symmetrically arranged
halfs of the SSO potential for $q > 0$. The structure of singularities at $q=0$ follows equally from
the discussion in section \ref{sec:pot}.

\begin{figure}
\hspace*{2cm}
\includegraphics[width=20cm,height=8cm]{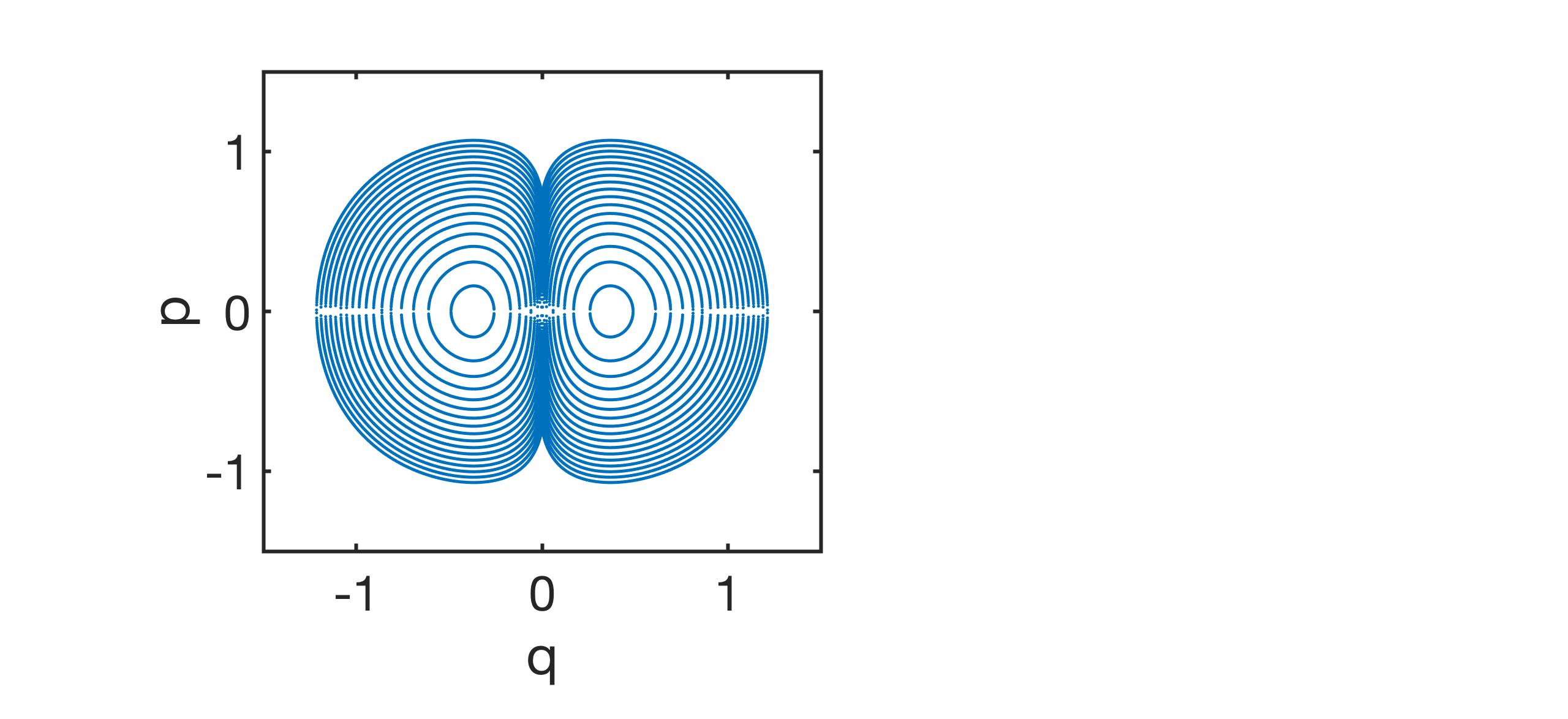} \vspace*{-1cm}
\caption{\label{fig:fig9} The phase space of the variant of the SSO with the 
potential $V(q)=|q|^{|q|}$.}
\end{figure}

\begin{figure}
\hspace*{2cm}
\includegraphics[width=20cm,height=8cm]{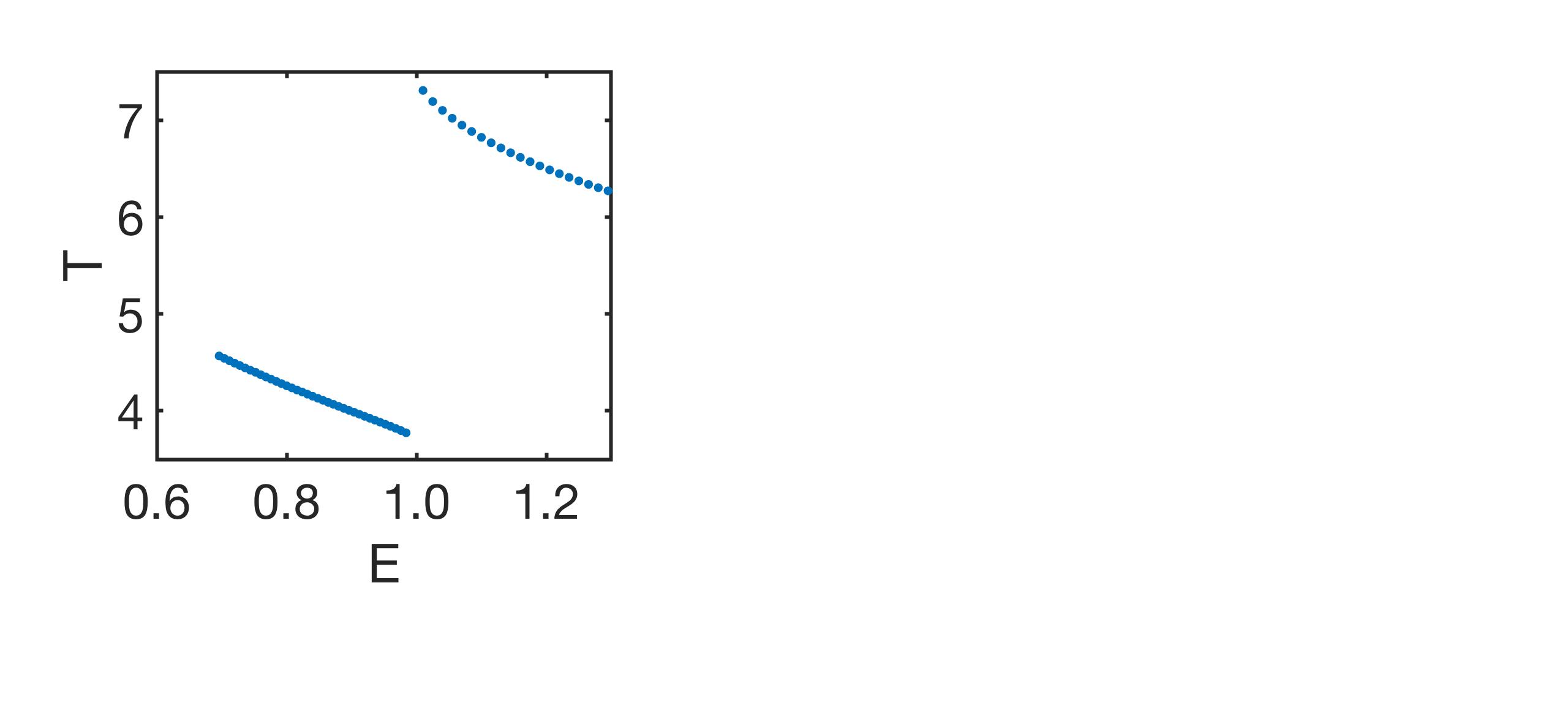} \vspace*{-1cm}
\caption{\label{fig:fig10} The period $T(E)$ as a function of the energy for the SVO
shows two branches with a decreasing linear and nonlinear behaviour, respectively.}
\end{figure}

The phase space of the SVO is shown in Figure \ref{fig:fig9}. It shows an agglomeration of
phase space curves at $q=0$ as the SSO does. Finally Figure \ref{fig:fig10} shows the period
of the SVO as a function of the energy. Two branches are visible. (i) A linearly decreasing branch for energies
below the value one which is identical to the one of the SSO and (ii) a nonlinearly decreasing branch
for energies above the maximum of the double well. The two branches are separated by a gap which simply 
stems from the opening of the phase space from the individual wells to the complete double well.

\section{Conclusions and Outlook} \label{sec:con}

Oscillators represent a key ingredient of our modelling of more complex systems and a starting-point for
the bottom-up understanding of fundamental dynamical processes in nature. While the physics of harmonic
and anharmonic coupled oscillators has been explored in great detail in the literature with a plethora
of resulting nonlinear phenomena we have pursued here a different approach and designed an oscillator which, in
a certain sense, puts the nonlinearity to the extreme: as for normal oscillators a simple power potential is employed, but
now both the basis and the exponent of it are coordinate dependent and determine therefore the overall
dynamics. In spite of its overall simplicity this superexponential self-interacting oscillator bears
a rich behaviour. While standard oscillators $V \propto q^{2n}$ of an equal simplicity are purely confining 
the SSO with its coordinate dependent $V \propto {|q|^q}$ shows both confined motion and scattering 
trajectories. The spatial dependence of the SSO potential leads to an extreme variation of the
nonlinearity ranging from a strongly confining to long-range decaying behaviour. The confining part
of the SSO consists of a strongly asymmetric potential well which leads to a bimodal behaviour of the
dynamics of the oscillator due to the existence of a transition point. At this transition point there
is a characteristic behaviour of the SSO potential with a hierarchy of logarithmic and power
law singularities of increasing order with increasing higher derivative. As a consequence the
period of the oscillator exhibits a crossover between two qualitatively different behaviours. For
energies below the transition point the period decreases (approximately) linearly with the energy 
whereas above it a strongly nonlinear dependence is observed finally leading to a monotonous increase
of the period. While already the linear decrease is a peculiar property of the SSO, the combination
of both behaviours below and above the transition point is remarkable for a system as simple as the
SSO is. The fingerprint of this transition in phase space is an agglomeration of the phase space curves.

Future directions of investigation could be the generalization of the SSO to higher dimensional 
space or the inclusion of dissipation and driving.  
Together with the scattering motion that is part of the phase space of the SSO, our analysis raises the
perspective that the SSO could be a fundamental building block for more complex few- and many-body oscillator systems
following the route which standard oscillators have taken. Certainly, there is major differences between
the standard (an-)harmonic oscillators and the SSO, but the rich complexity of the behaviour the SSO achieved
already with a single degree of freedom is very promising towards coupling few or many of them, which could
be performed in several ways. Finally we note that the extreme nonlinear and asymmetric behaviour of the SSO points
in the direction that it could be an effective description of a complex (or even biological) many
degree of freedom system which, again on a higher hierarchical level, could be interconnected.

\section{Acknowledgments}

This work has been in part performed during a visit to the Institute for Theoretical Atomic, Molecular and Optical Physics (ITAMP)
at the Harvard Smithsonian Center for Astrophysics in Cambridge, Boston, whose hospitality is gratefully acknowledged.
The author thanks B. Liebchen for a careful reading of the manuscript and valuable comments. Illuminating discussions
with F.K. Diakonos are equally acknowledged.

\end{document}